\documentclass[reprint,superscriptaddress,amsmath,twocolumn,secnumarabic,amssymb, nobibnotes, aps, prl]{revtex4-2}

\usepackage{hyperref}
\usepackage{amsmath,bm,lipsum,mathtools}
\usepackage{amssymb}
\usepackage{graphicx}
\usepackage{tensor}

\usepackage{float}
\usepackage{svg}

\hypersetup{hypertex=true,colorlinks=true,linkcolor=blue,anchorcolor=blue,citecolor=blue,urlcolor=blue}

\begin{document}
\title{\textcolor{red}{Energization of Proton via Beam-Driven Ion Bernstein Waves in $\mathbf{p\prescript{11}{}{B}}$ Plasmas}}
\author{Yangchun Liu$^\dagger$}
\affiliation{Institute for Fusion Theory and Simulation, School of Physics, Zhejiang University, Hangzhou 310058, China}
\author{Hairong Huang$^\dagger$}
\affiliation{Hebei Key Laboratory of Compact Fusion, and ENN Science and Technology Development Co., Ltd, Langfang 065001, China}
\author{Dong Wu}
\email{dwu.phys@sjtu.edu.cn}
\affiliation{State Key Laboratory of Dark Matter Physics, Key Laboratory for Laser Plasmas (MoE), School of Physics and Astronomy, Shanghai Jiao Tong University, Shanghai 200240, China}
\author{Tianxing Hu}
\affiliation{State Key Laboratory of Dark Matter Physics, Key Laboratory for Laser Plasmas (MoE), School of Physics and Astronomy, Shanghai Jiao Tong University, Shanghai 200240, China}
\author{Huasheng Xie}
\affiliation{Hebei Key Laboratory of Compact Fusion, and ENN Science and Technology Development Co., Ltd, Langfang 065001, China}
\author{Bing Liu}
\email{liubingw@enn.cn}
\affiliation{Hebei Key Laboratory of Compact Fusion, and ENN Science and Technology Development Co., Ltd, Langfang 065001, China}
\author{Zhengmao Sheng}
\affiliation{Institute for Fusion Theory and Simulation, School of Physics, Zhejiang University, Hangzhou 310058, China}
\author{Jiaqi Dong}
\affiliation{Southwestern Institute of Physics, Chengdu, 610041, China}
\author{Yueng-Kay Martin Peng}
\affiliation{Hebei Key Laboratory of Compact Fusion, and ENN Science and Technology Development Co., Ltd, Langfang 065001, China}

\thanks{$^\dagger$These authors contributed equally to this work.}

%\date{}

\begin{abstract}
	Energizing background ions plays a pivotal role in all forms of thermal nuclear fusion, as it can increase the fusion reaction rate without affecting the overall mechanical equilibrium. This is particularly critical for $\mathbf{p\prescript{11}{}{B}}$ fusion due to its exceptionally high operating temperature and substantial energy losses from bremsstrahlung radiation. Here, we report a nonlinear mechanism that efficiently transfers the energy of injected heating beams to background protons in $\mathbf{p\prescript{11}{}{B}}$ mixed plasmas, via fully kinetic Particle-In-Cell (PIC) simulations. When a proton neutral beam is injected into $\mathbf{p\prescript{11}{}{B}}$ plasmas, it triggers the excitation of ion Bernstein waves (IBWs) at harmonics of the proton cyclotron frequency. In the initial linear stage, the energy channels to background electrons and protons might be comparable, consistent with theoretical model for the energy transfer. However, in the latter nonlinear stage, the dominant channel transfers to background protons, generating a non-Maxwellian population of energetic protons. This transition is driven by a nonlinear spectral cascade of IBWs toward lower frequencies and longer wavelengths, which strengthens wave proton coupling while suppressing wave electron coupling.
\end{abstract}

\maketitle

	\textit{Introduction}---In thermonuclear fusion plasmas, the fusion power output is intrinsically controlled by the kinetic properties of the ion population, as the reaction rate depends sensitively on the ion velocity distribution through the velocity averaged fusion reactivity $ \left< \sigma v \right> $ \cite{Lawson1957PPSB,wesson2011tokamaks}. Since the fusion cross section rises steeply with increasing ion energy in the relevant energy range, targeted enhancement of ion energies can lead to a substantial increase in fusion reactivity even when the macroscopic plasma parameters remain unchanged. This characteristic enables fusion performance to be optimized by modifying microscopic particle dynamics rather than altering global equilibrium conditions such as pressure balance or magnetic configuration.	This principle underpins modern auxiliary heating techniques such as neutral beam injection and radio frequency wave heating which are widely used to selectively energize ions and tailor their velocity distributions \cite{Duesing1987FT,litvak1977NF}. Beyond purely thermal effects, recent theoretical and numerical studies have demonstrated that non-Maxwellian features, such as energetic ion tails or anisotropic temperature distributions, can further enhance fusion reactivity under reactor relevant conditions, highlighting the potential of kinetic tailoring as a pathway toward higher fusion gain \cite{Kong2024PPCF,Nicks2021NF,Magee2019NP,Squarer2024PRE}.
	\par
	This need for precise ion energization becomes especially critical in advanced fuel cycles such as $\mathbf{p\prescript{11}{}{B}}$ fusion. Although $\mathbf{p\prescript{11}{}{B}}$ fusion offers advantages including abundant fuel and aneutronic charged particle products, it requires ion temperatures roughly an order of magnitude higher than those of $\mathbf{DT}$ fusion. The concomitant increase in bremsstrahlung radiation historically posed a severe power balance challenge, leading early analyses to deem the reaction infeasible \cite{Nevins1998JFE,Rider1997PoP}. Revised cross section data \cite{Sikora2016JFE} and improved power balance models have since identified accessible regimes, particularly when hot ion operation model ($T_i/T_e>2$) is sustained to suppress radiative losses \cite{Cai2022FST,Liu2025PoP}. Recent experiments on JET and EAST have demonstrated such hot ion operation modes, confirming their experimental viability \cite{Mantica2024NF,Seo2021NF}.
	\par
\begin{figure}
		\centering
		\includegraphics[scale=0.5]{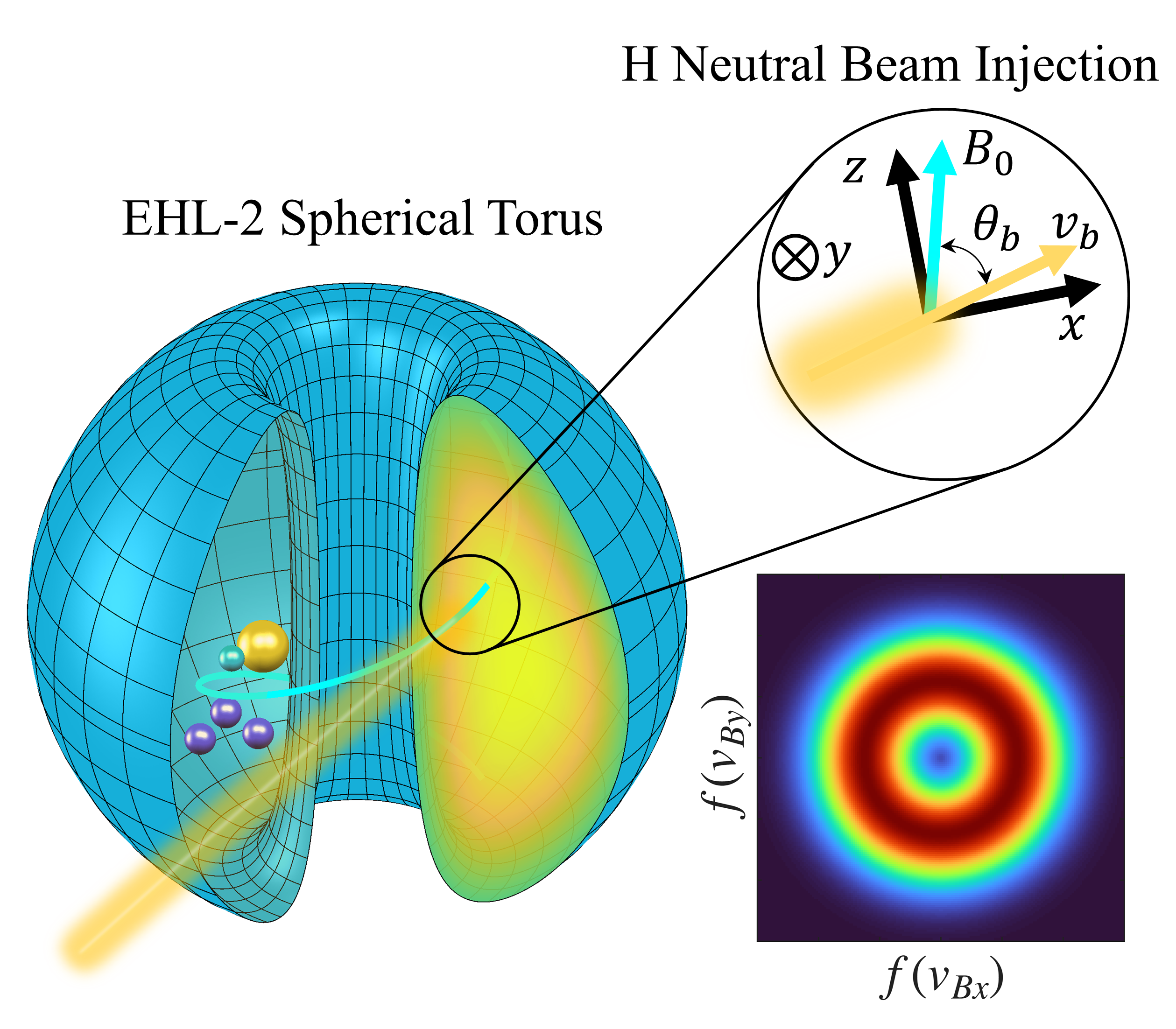}
		\caption{Schematic of neutral beam injection into $\mathbf{p\prescript{11}{}{B}}$ plasma. The beam is injected obliquely at an angle $\theta_b$ relative to the background magnetic field $B_0$ and follows an annular distribution perpendicular to the field.}\label{fig1}
\end{figure}
	Pushing the frontier of ion energization, recent research focuses on manipulating energy transfer pathways between fusion products and fuel ions. A prominent example is $\alpha$-channeling, which uses plasma waves to direct fusion born $\alpha$‑particle energy selectively into fuel ions, bypassing electron heating and thereby mitigating bremsstrahlung losses \cite{Fisch1992PRL, Ochs2022PRE}. Concurrently, wave particle interaction schemes such as beam‑driven IBWs have been shown in simulations and experiments to selectively energize specific ion species, further illustrating the potential of externally driven, nonequilibrium distributions to enhance fusion reactivity \cite{Magee2019NP, Nicks2021NF}. These advances underscore ion energization not merely as a heating method, but as a essential tool for realizing advanced fuel fusion systems. 
\par
\begin{figure}
	\centering
	\includegraphics[scale=0.6]{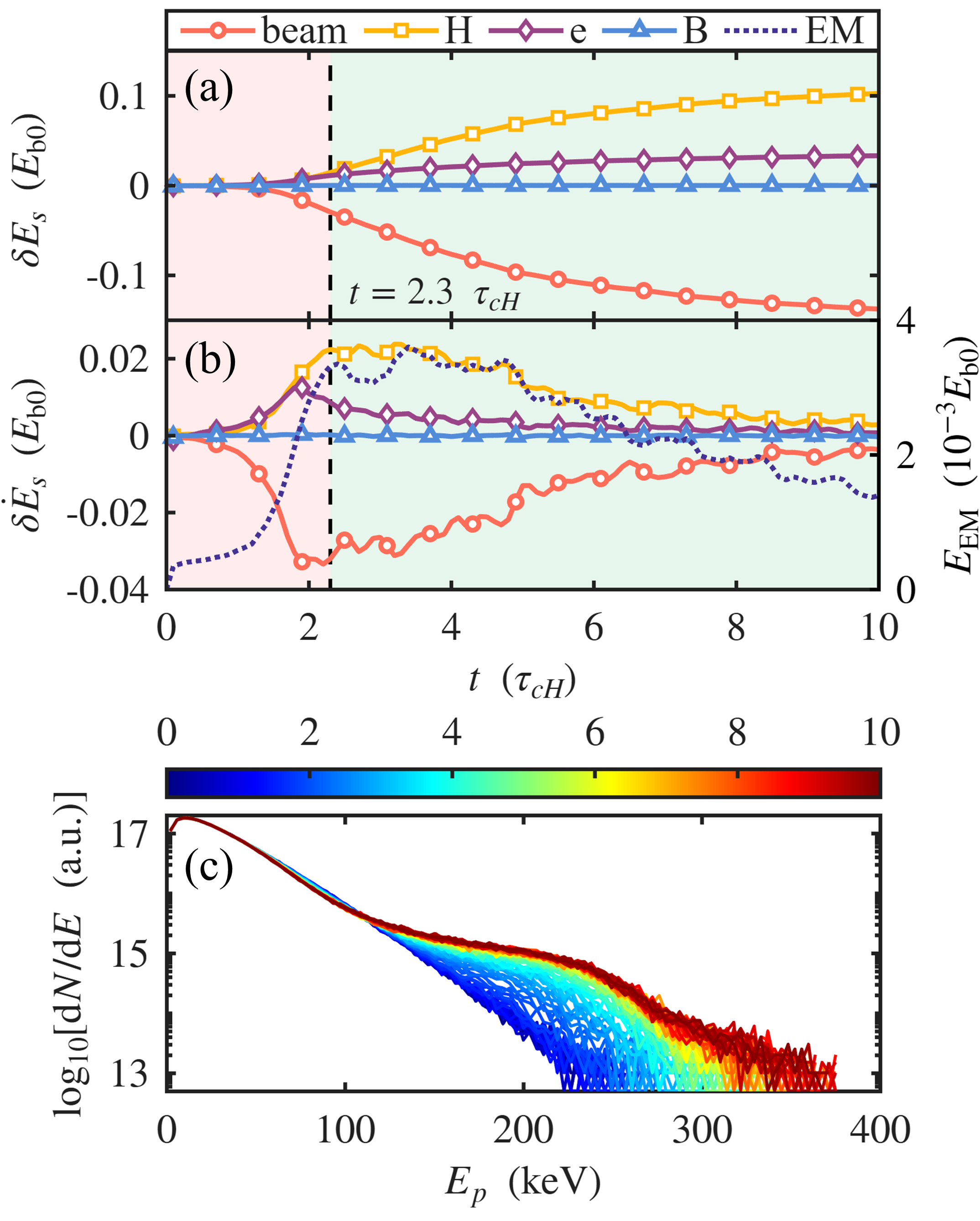}
	\caption{(a) Temporal evolution of the energy change of each particle species. (b) Energy transfer rate per unit initial beam energy and temporal evolution of the electromagnetic field energy. (c) Temporal evolution of the background protons energy spectra (colored by time).}\label{fig2}
\end{figure}
\par
	In this letter, we, for the first time, propose and explore the collisionless energy transfer channel within $\mathbf{p\prescript{11}{}{B}}$ mixed plasmas. Our focus is on the interaction between proton neutral beams and $\mathbf{p\prescript{11}{}{B}}$ plasmas at the outer midplane edge of magnetic confinement devices. \textit{Ab initio} fully kinetic one–spatial and three–velocity dimension (1D3V) simulations demonstrate that injected neutral beams are capable of exciting unstable IBWs on a timescale significantly shorter than the collision timescale. During the linear stage, the majority of the energy originating from neutral beams is transferred to background electrons, while boron ions gain almost no energy. This observation is in line with the predictions of linear theory. However, as the system evolves over an extended period in the nonlinear stage, the dominant energy transfer channel shifts to background protons, resulting in the generation of a non–Maxwellian population of energetic protons. This transition is a nonlinear spectral cascade toward longer wavelengths and lower frequencies, which enhances wave proton coupling while weakening interaction with electrons. This mechanism provides confirmation of a potential pathway that energize background ions in a collisionless way.
\par
\begin{figure}
	\centering
	\includegraphics[scale=0.6]{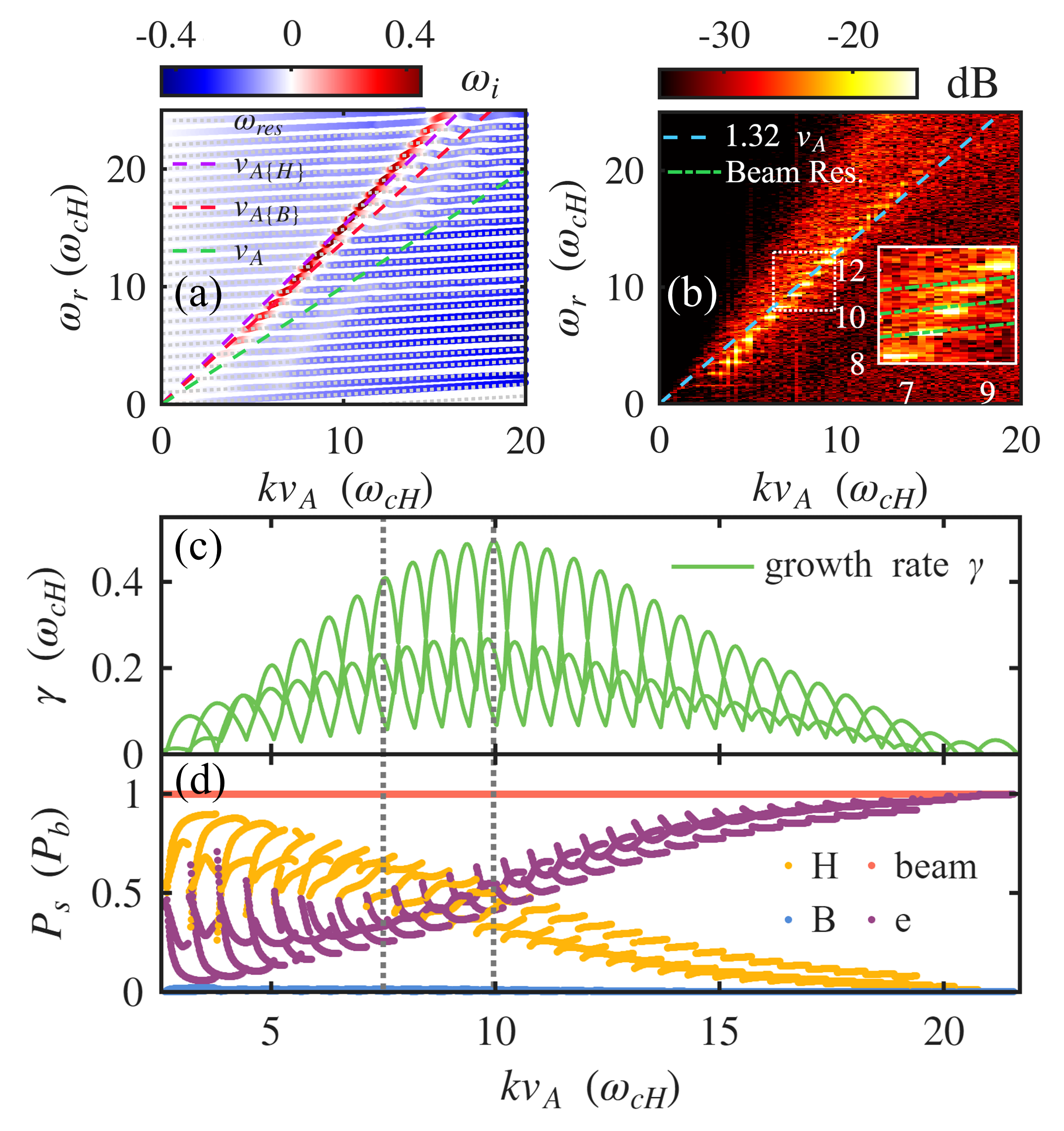}
	\caption{(a) Dispersion relation from PDRK/BO. Purple and red dashed lines denote the Alfvén velocities of background boron ions and protons, given by $ v_{A\{\text{H,B}\}}=c/\sqrt{\omega _{p\left\{ \text{H,B} \right\}}/\omega _{c\left\{ \text{H,B} \right\}}}$. Green dashed line represents the total Alfvén velocity $v_A=c/\sqrt{\sum_s{\left( \omega _{ps}/\omega _{cs} \right)}}$, while the gray dotted line indicates the beam resonance condition. (b) The dispersion relation for z-component of the fluctuating magnetic field $\delta B_z$. The blue dashed and green dash dotted lines represent $\omega=1.32kv_A$ and the beam resonance condition, respectively. (c) The growth rate versus wave vector. (d) Energy transfer ratio normalized to beam energy. Gray dashed lines indicate $7.5kv_A/\omega _{c\text{H}}$ and $9.96kv_A/\omega _{c\text{H}}$, respectively.}
	\label{fig3}
\end{figure}
	\textit{Setup}---ENN Science and Technology Development Co., Ltd. (ENN) is advancing $\mathbf{p\prescript{11}{}{B}}$ fusion through a strategic roadmap centered on the spherical torus (ST) concept. As a key step in this program, the next generation device EHL‑2 has been designed to validate high‑performance ST operation modes, including the achievement of hot ion operation modes with $T_i/T_e>2$ and the demonstration of measurable $\alpha$ particle yield from $\mathbf{p\prescript{11}{}{B}}$ reactions \cite{Liu2024PoP,Liang2025PST}. To support the physics basis of the EHL‑2 design, a set of kinetic simulations has been performed by using the \textit{Ab initio} high-order implicit PIC code LAPINS \cite{Wu2019PRE,SupplementaryMaterial}, with the results validated against those obtained from the open source full kinetic PIC code EPOCH \cite{Arber2015PPCF}. In the present work, we focus on the energy deposition of a proton neutral beam near the outer midplane edge of EHL-2. The neutral beam is injected at  $\theta_b=60^\circ$  relative to the magnetic field in the $\mathbf{p\prescript{11}{}{B}}$, as illustrated in Fig.~\ref{fig1}. Fast ions are described by a ring beam in velocity space
	\begin{equation}
		f\propto \exp \left[ -\frac{\left( v_{\parallel}-v_{b\parallel} \right) ^2}{v_{tb}^{2}} \right ] \exp \left[ -\frac{\left(v_{\bot}-v_{b\bot} \right)^2}{2v_{tb}^{2}} \right],
	\end{equation}
	where $v_\parallel$ and $v_\perp$ denote velocity components parallel and perpendicular to the magnetic field, $v_{b\parallel}$ and $v_{b\perp}$ are constants that define the average drift speed, and $v_{tb}$ controls the thermal spread. The plasma is modeled as a locally homogeneous mixture of proton and boron with an injected proton neutral beam in a uniform magnetic field of strength $B_0=2\ \mathrm{T}$, oriented at $\theta_k= 85^\circ$ relative to the simulation domain. The boron density is set to $n_\mathrm{B}=9.2\times10^{18}\ \mathrm{m^{-3}}$, the background proton density to $n_\mathrm{H}=9n_\mathrm{B}$, and the beam density to $n_\mathrm{b}=8\times10^{18}\ \mathrm{m^{-3}}$. The beam energy is $E_\mathrm{b}=200\ \mathrm{keV}$, giving a total speed $v_\mathrm{b}=\sqrt{2E_\mathrm{b}/m_\mathrm{H}}$. Its parallel and perpendicular components are defined as $v_{b\parallel}=v_b\cos{\theta_b}$ and $v_{b\perp}=v_b\sin{\theta_b}$. Quasineutrality determines the electron density. This domain is discretized into $6820$ grid, smaller than the NBI ion gyroradius to properly resolve cyclotron motion. Macroparticle numbers for proton and boron ions are assigned proportionally to their densities, with 128 boron macroparticles per cell. This local 1D3V approximation, which has been successfully applied in previous studies of fast ion driven instabilities in C‑2U \cite{Nicks2021NF,Magee2019NP}, KSTAR \cite{Chapman2017NF,Chapman2018NF}, and LHD \cite{Reman2019NF}, retains the essential physics while neglecting toroidal effects that are secondary for the mechanisms under investigation.
\par
	\textit{Simulation Results}---Fully kinetic simulations demonstrate an efficient collisionless channel for transferring neutral beam energy to background ions. As shown in the  Fig.~\ref{fig2}(a), a significant fraction of the beam energy is redistributed within only a few proton cyclotron periods. Immediately after beam injection, electromagnetic fluctuations grow exponentially, signaling the onset of beam-driven collective instabilities.  During this early evolution, the cumulative energy gains of electrons and background proton are comparable, whereas boron ions exhibit negligible energization. As the system evolves further, a pronounced bifurcation in the energy partition emerges. Protons continue to gain energy at an accelerated rate and gain over $10\%$ of the beam energy, while electron energization saturates. This later behavior indicates that the dominant energy channel shifts from electron dominated damping to preferential ion heating. 
\par
	Fig.~\ref{fig2}(b) presents the temporal evolution of particle energy transfer rates and electromagnetic field energy. The temporal evolution of energy transfer rates and electromagnetic field energy provides direct insight into the underlying mechanism. The beam initially transfers its free energy predominantly to electromagnetic fields, whose energy growth closely follows the instability growth rate. These fields subsequently redistribute energy to different particle species through wave–particle interactions. Notably, the electromagnetic field energy peaks and then rapidly declines, coincident with a sharp increase in proton energy, demonstrating that the fields act as an intermediate reservoir rather than a terminal sink. Besides, the temporal evolution of the background protons energy spectra in Fig.~\ref{fig2}(c) show a non-Maxwellian population of energetic ions which demonstrate there may be different mechanisms of energy transfer in both linear and nonlinear stages.
\par 
\begin{figure}
	\centering
	\includegraphics[scale=0.6]{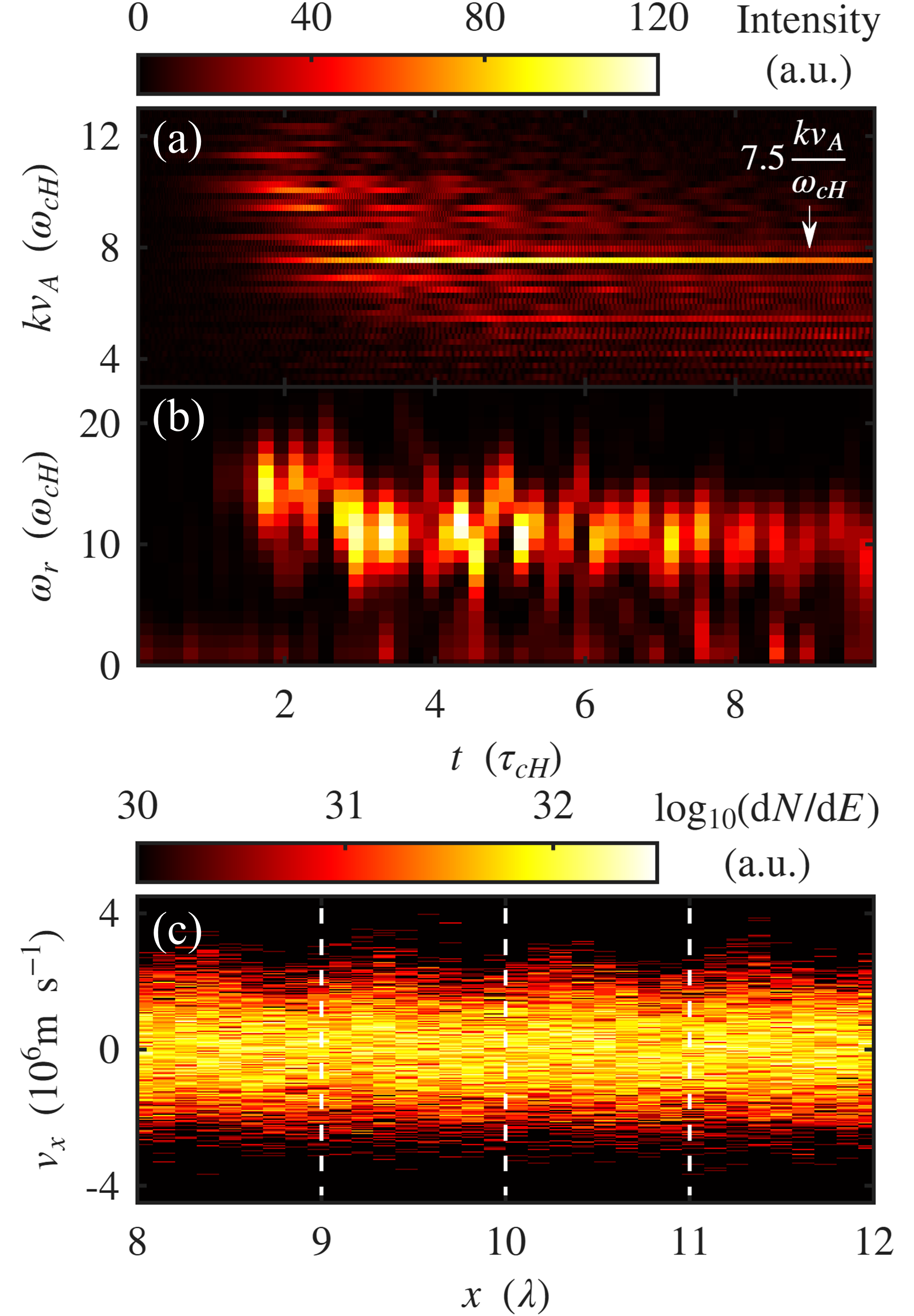}	
	\caption{(a) and (b) are temporal evolution of $\boldsymbol{k}$ and $\omega$. (c) A snapshot of the the phase space of background protons.}
	\label{fig4}
\end{figure}
	\textit{Linear analyse}---To interpret the simulation results, we analyze wave dynamics and energy transfer rates using a linear kinetic framework. The system is described by the Vlasov and Maxwell's equations, which in Fourier space yield the wave equation for a collisionless plasma
	\begin{equation}
		\boldsymbol{k}\times \left( \boldsymbol{k}\times \hat{\boldsymbol{E}} \right) +\frac{\omega ^2}{c^2}{\boldsymbol{\epsilon }}\cdot \hat{\boldsymbol{E}}=0,
		\label{DSequation}
	\end{equation} 
where
	\begin{equation}
		{\boldsymbol{\epsilon }}=\frac{4\pi i}{\omega}{\boldsymbol{\sigma }} +{\boldsymbol{I}},
	\end{equation}
$\hat{\boldsymbol{E}}$ is the wave electric field in Fourier space, ${\boldsymbol{\epsilon }} $ is the dielectric tensor, ${\boldsymbol{\sigma }} $ is the conductivity tensor, $\omega$ and $\boldsymbol{k}$ are the wave frequency and wave vector, and ${\boldsymbol{I}} $ represents the identity matrix. The conductivity tensor can be divided among the contributions from each species $s$:
	\begin{equation}
		{\boldsymbol{\sigma }}=i\sum_s{\frac{q_{s}^{2}n_s}{m_s\omega _{cs}}\sum_n{\int_{-\infty}^{\infty}{\int_0^{\infty}{\frac{2\pi \boldsymbol{P}_{sn}v_{\bot}\text{d}v_{\bot}\text{d}v_{\parallel}}{\left( k_{\parallel}v_{\parallel}-\omega \right) /\omega _{cs}+n}}}}}
	\end{equation}
where $q_s$, $m_s$ and $n_s$ are the species charge, species mass and species density, respectiveley. $\omega_{cs}$ is species cyclotron frequency. $k_\parallel$ and $k_\perp$ denote wave vector components parallel and perpendicular to the magnetic field. $\boldsymbol{P}_{sn}$ is a matrix related to distribution function $f$, and its specific expression is referred to \cite{Bhattacharjee2005}. 
\par
The plasma wave eigenmodes correspond to solutions of Eq.~\ref{DSequation}. Once the wave electric field and magnetic field fluctuations are obtained, the wave electromagnetic energy can be wrote as \cite{Liu2021ApJ,Zhao2022ApJ,Howes2017JPLPH,Klein2017JPLPH,Klein2019ApJ,Klein2020JPLPH}
	\begin{equation}
		W_{\mathrm{EB}}=\frac{1}{4}\left( \epsilon _0\left| \hat{\boldsymbol{E}} \right|^2+\frac{1}{\mu _0}\left| \hat{\boldsymbol{B}} \right|^2 \right),
	\end{equation}
	where $\epsilon_0$ and $\mu_0$ are the permittivity and the permeability of free space, $\hat{\boldsymbol{B}}$ are the wave magnetic field in Fourier space. The energy transfer rate between the waves and particles can be quantified by
	\begin{equation}
		\Gamma _s=\frac{1}{4}\left( \hat{\boldsymbol{E}}\cdot \hat{\boldsymbol{J}}_{s}^{*}+\hat{\boldsymbol{E}}^*\cdot \hat{\boldsymbol{J}}_s \right),
	\end{equation}
	which denotes the energy transfer per unit of time and per unit of volume, with $\boldsymbol{J}_s=\boldsymbol{\sigma}_s\cdot \boldsymbol{\hat{E}}$ corresponding to the species current. The energy transfer between the waves and particles can be express by
	\begin{equation}
		P_s=\frac{\Gamma _s}{W_{\mathrm{EB}}},
	\end{equation}
	which means the energy transfer emission per unit of time, per unit of volume, and per unit of wave electromagnetic energy. In this work, the dispersion relation and energy transfer ratio are computed numerically using the kinetic dispersion solver BO/PDRK \cite{Xie2016PST,Xie2019CPC,Xie2025PoP}, which implements the above theoretical framework and has been benchmarked in previous linear analyses of beam-driven instabilities \cite{Liu2021ApJ,Zhao2022ApJ}.
\par
	The linear stage of the instability is characterized in the Fig.~\ref{fig3}. A spatiotemporal Fourier analysis of the fluctuating magnetic fields shows that the unstable spectrum is dominated by IBWs excited near harmonics of the proton cyclotron frequency. These modes satisfy the cyclotron resonance condition $\omega =k_{\parallel}v_{b\parallel}\pm n\omega _{c\text{H}}$ and propagate with phase velocities close to the Alfvén speed of proton. Kinetic dispersion relationship calculations reproduce both the unstable frequency bands and the locations of maximum growth, confirming that the observed waves correspond to the fastest growing linear eigenmodes. Within this linear regime, the energy transfer rate analysis predicts that the beam supplies energy to the waves, which are subsequently damped by electrons and protons with comparable strength. This prediction is borne out by the simulations, with electron and ion energy gains track each other closely during the first few cyclotron periods. Boron ions, by contrast, remain weakly coupled because the wave frequencies lie far above their cyclotron frequency. At this stage, the system dynamics are well described by linear theory, and the energy partition follows directly from the relative wave particle coupling strengths encoded in Fig.~\ref{fig3}(d).
\par
\begin{figure}
	\centering
	\includegraphics[scale=0.6]{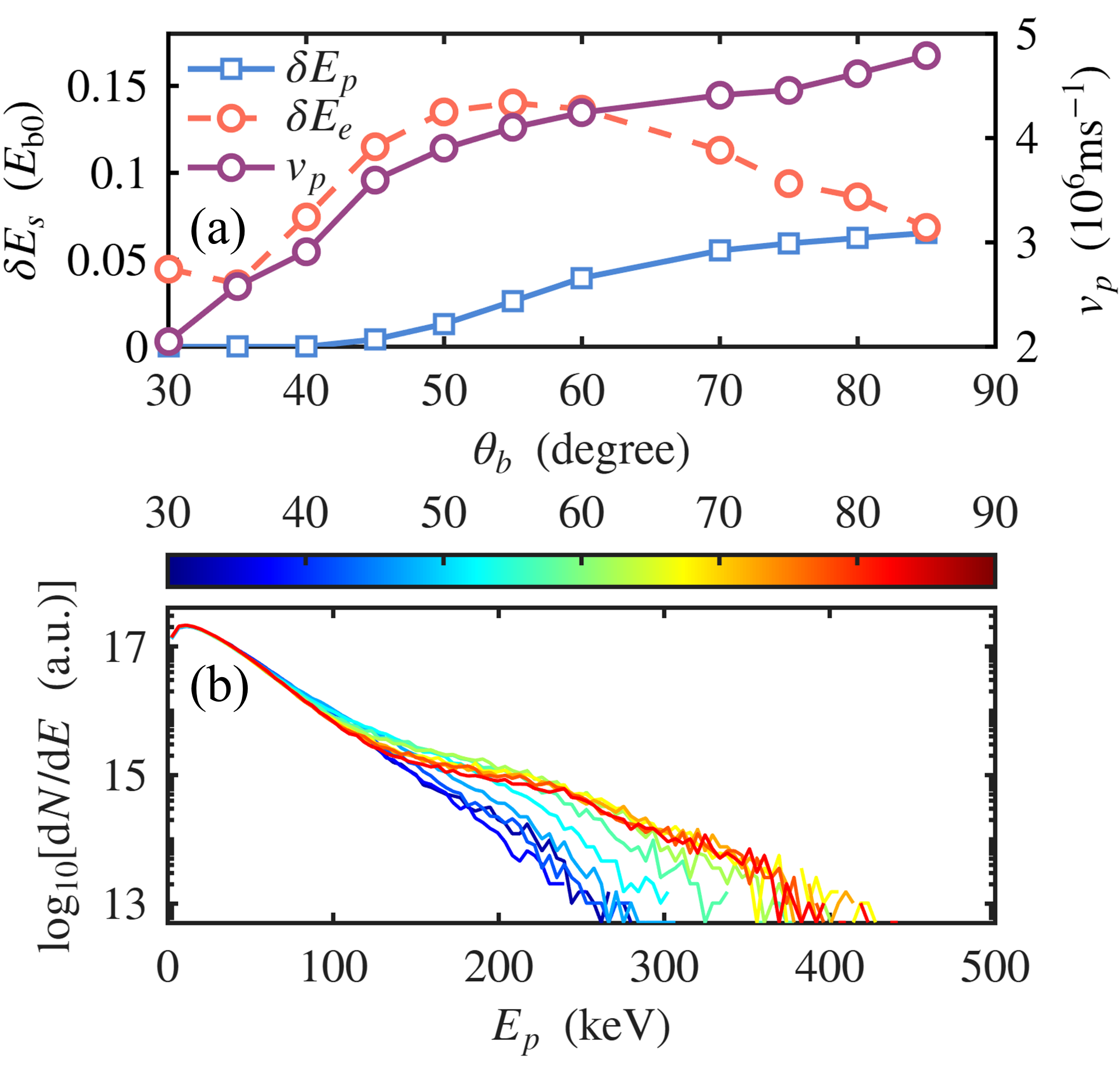}	
	\caption{(a) Electron and ion energy transfer rates and phase velocity of excited IBWs as functions of $\theta_b$ at nonlinear saturation. (b) Background proton energy spectra for different $\theta_b$ at nonlinear saturation.}
	\label{fig5}
\end{figure}
	\textit{Nonlinear analyse}---As IBWs enters the nonlinear stage, the system departs from the initial linear predictions and undergoes a qualitative transition in its energy partition. As summarized in Fig.~\ref{fig4}(a) and (b), this phase is primarily characterized by a nonlinear spectral cascade where the dominant IBWs shift systematically toward lower frequencies and longer wavelengths. This spectral migration fundamentally alters the wave particle interaction landscape by weakening the resonant interactions with background electrons while simultaneously enhancing the resonance interaction with background protons. Despite the inherent nonlinearity of this spectral evolution, the energy transfer pathways can still be analyzed using the linear kinetic framework. Specifically, the energy transfer rate model remains applicable for quantifying the power exchange between the shifted wave spectrum and the plasma species which are shown in Fig.~\ref{fig3}(c) and (d). By utilizing the numerical energy transfer ratio defined in the linear theory, we can demonstrate that the shifting spectral properties redistribute the damping pathways in favor of preferential ion heating. This confirms that the linear model provides a robust diagnostic tool for the energy partition even after the onset of nonlinear spectral modification.
\par
The sustained ion wave coupling drives a rapid divergence between proton and electron energy gains. As shown in Fig.~\ref{fig4}(c), phase space diagnostics reveal that protons become periodically trapped and accelerated by the wave electric fields—a signature behavior of wakefield-like acceleration \cite{Nicks2021NF}. This mechanism generates a pronounced non-Maxwellian high-energy tail in the proton distribution as shown in Fig.~\ref{fig1}(c). Throughout this evolution, boron ions remain largely unaffected because the modified wave frequencies continue to lie far above the boron cyclotron frequency, maintaining their weak coupling to the energy channel. Consequently, the nonlinear stage enables an efficient collisionless transfer of neutral beam energy into the fuel ions, providing a viable pathway for effective energization background ions.
\par
To further investigate the optimization of this mechanism, we performed a parameter scan regarding the neutral beam injection angle	at the nonlinear saturation stage. As illustrated in Fig.~\ref{fig5}(a), the phase velocity of IBWs decreases monotonically as the injection angle is reduced. Consequently, the energy transfer rate to electrons exhibits a consistent decline with smaller angles. This reduction occurs because the phase velocity remains significantly lower than the electron thermal velocity, and a decreasing phase velocity further weakens the coupling between the waves and the electrons. In contrast, the energy transfer to protons displays a non-monotonic trend that initially increases and then decreases with rising injection angles. This behavior arises from a competition between the acceleration limit and the trapping efficiency. The cutoff velocity for accelerated background ions is approximately twice the phase velocity, meaning higher phase velocities theoretically allow for greater energy gains. However, a larger discrepancy between the phase velocity and the background ion thermal velocity reduces the number of ions that can be effectively trapped. The evolution of the proton energy spectra presented in Fig. \ref{fig5}(b) corroborates this interplay, indicating that an optimal injection angle exists to maximize total ion energization by balancing the energetic tail extent with the population of accelerated particles.
\par
	In summary, the study demonstrates a nonlinear collisionless energy transfer process via beam-driven IBWs in $\mathbf{p\prescript{11}{}{B}}$ mixed plasmas. The linear stage is dominated by wave growth and comparable energy deposition into both background electrons and protons, consistent with kinetic theory. The nonlinear stage, however, is characterized by a spectral cascade toward lower frequencies and longer wavelengths, which enhances wave proton coupling and weakens wave electron coupling, promotes the formation of a non‑Maxwellian proton population, and ultimately channels most of the beam energy into protons. This transition fundamentally changes the energy partition and enables efficient ion heating without relying on collisional processes.
\par
	These findings highlight a promising mechanism for energizing background protons in $\mathbf{p\prescript{11}{}{B}}$ fusion plasmas. By channeling neutral beam energy preferentially into background protons, beam-driven IBWs may significantly enhance fusion reactivity while mitigating electron energy losses. Future work should explore multidimensional effects, parameter sensitivity, and experimental validation to assess the robustness of this mechanism under realistic reactor conditions.

This work was supported by the National Key R$\&$D Program of China (Grant NO. 2019YFE03050001), and the Energy iNNovation (ENN) Science and Technology Development Co. (Grant No. 2021HBQZYCSB006).

\bibliography{reference.bib}
\end{document}